\documentclass[conference]{IEEEtran}
\IEEEoverridecommandlockouts

\usepackage{cite}
\usepackage{amsmath,amssymb,amsfonts}
\usepackage{algorithmic}
\usepackage{graphicx}
\usepackage{textcomp}
\usepackage{xcolor}
\usepackage{bbm}
\usepackage{url}
\def\BibTeX{{\rm B\kern-.05em{\sc i\kern-.025em b}\kern-.08em
    T\kern-.1667em\lower.7ex\hbox{E}\kern-.125emX}}
\begin{document}

\title{Task- and Metric-Specific Signal Quality Indices for Medical Time Series
\thanks{This research is funded by the European Union’s Horizon Europe programme under the Grant Agreement no. 101137278. The source code is available at \protect\url{https://github.com/JH2k00/perturbation-sqi}.}
}

\author{\IEEEauthorblockN{Jad Haidamous, Christoph Hoog Antink}
\IEEEauthorblockA{\textit{KIS*MED - AI Systems in Medicine} \\
\textit{Technical University of Darmstadt}\\
Darmstadt, Germany \\
\{haidamous, hoogantink\}@kismed.tu-darmstadt.de}
}

\maketitle

\begin{abstract}
Medical time series such as electrocardiograms (ECGs) and photoplethysmograms (PPGs) are frequently affected by measurement artifacts due to challenging acquisition environments, such as in ambulances and during routine daily activities. Since automated algorithms for analyzing such signals increasingly inform clinically relevant decisions, identifying signal segments on which these algorithms may produce unreliable outputs is of critical importance. Signal quality indices (SQIs) are commonly used for this purpose. However, most existing SQIs are task agnostic and do not account for the specific algorithm and performance metric used downstream. In this work, we formalize signal quality as a task- and metric-dependent concept and propose a perturbation-based SQI (pSQI) that aims to detect an algorithm's performance degradation on an input signal with respect to a metric. The pSQI is defined as the worst-case value of the performance metric under an additive, colored Gaussian noise perturbation with a lower-bounded signal-to-noise ratio. We introduce formal requirements for task- and metric-specific SQIs, including monotonicity of the metric in expectation and maximal separation under thresholding. Experiments on R-peak detection and atrial fibrillation classification benchmarks demonstrate that the proposed pSQI consistently outperforms existing feature- and deep learning-based SQIs in identifying unreliable inputs without requiring training.
\end{abstract}

\begin{IEEEkeywords}
Signal Quality Index, ECG, PPG, R-Peak Detection, Atrial Fibrillation Classification.
\end{IEEEkeywords}

\section{Introduction}
The widespread adoption of wearable and mobile health monitoring devices has led to an unprecedented increase in the availability of physiological time-series such as electrocardiograms (ECGs) and photoplethysmograms (PPGs) recorded during routine daily activities~\cite{tan2021incentia11k, torres2020multi}. These signals are particularly susceptible to corruptions such as motion artifacts, intermittent sensor contact loss, and environmental interference~\cite{clifford2012wireless}. Large fractions of the recorded data may therefore be unreliable or unsuitable for further downstream analysis~\cite{orphanidou2014signal}. In this setting, signal quality indices (SQIs) have become a critical component of medical time-series processing pipelines in recent years~\cite{orphanidou2014signal}. SQIs are commonly used to identify and reject corrupted signal segments~\cite{clifford2011signal, orphanidou2014signal}, or are integrated into algorithms to help deal with noisy inputs~\cite{nemati2010data}. A wide variety of SQIs have been proposed in the literature, ranging from statistical measures, such as signal-to-noise ratios and spectral heuristics~\cite{clifford2011signal, nemati2010data}, to morphological descriptors, such as the correlation between beats and the agreement rate between peak detectors~\cite{clifford2011signal,orphanidou2014signal}. 

Despite their widespread use, most existing SQIs are task-agnostic: They are designed to capture generic signal properties, largely independent of the downstream algorithm~\cite{clifford2011signal}. In practice, however, signal quality is not an intrinsic property of the signal alone. A signal that is perfectly suitable for one algorithm, such as R-peak detection, may be unusable for another, such as atrial fibrillation (AF) classification. Recent approaches therefore define and develop SQIs relative to the downstream task with the help of expert knowledge or labeled task-specific datasets~\cite{orphanidou2014signal, syversen2025framework}. The same signal may however also be acceptable under a tolerant metric while being unacceptable under a strict one for the same task. This work therefore argues that signal quality should be defined relative to not only an algorithm, but a metric as well.

Based on these principles, we formalize realistic requirements for task- and metric-specific SQIs in Section~\ref{sec:formalizing_sqis}. We then introduce a novel and \textbf{label-free} task- and metric-specific perturbation-based SQI (pSQI) in Section~\ref{sec:psqi}, and empirically compare its performance against other SQIs on multiple tasks and modalities in Section~\ref{sec:empirical_evaluations}. We conclude in Section~\ref{sec:conclusion}.

\section{Formalizing Signal Quality Indices}
\label{sec:formalizing_sqis}

Let $\mathcal{X} \subset \mathbb{R}^{N}$ denote the set of univariate time-series signals with $N$ samples. An input signal is denoted by $x \in \mathcal{X}$. We consider a fixed algorithm $f : \mathcal{X} \rightarrow \mathcal{Y}$ which maps an input time-series to an output in an arbitrary output set $\mathcal{Y}$. The output set $\mathcal{Y}$ may represent continuous values (e.g.\ regression outputs), discrete labels (e.g.\ class predictions), or sets of detected events (e.g.\ R-peaks). Let $h : \mathcal{Y} \times \mathcal{Y} \rightarrow \mathbb{R}$ denote a performance metric that assigns a real-valued score to a pair of outputs, where higher values indicate better performance. Given a ground-truth output $y \in \mathcal{Y}$ and a prediction $\hat{y} = f(x)$, the value $h(\hat{y}, y)$ quantifies the quality of the algorithm's output on input $x$. The metric may be continuous or discrete.

An SQI is defined as a scalar-valued function $q : \mathcal{X} \rightarrow \mathbb{R}$ where higher values of $q(x)$ indicate higher signal quality.

\subsection{Requirements for Task- and Metric-Specific SQIs}

We now formalize two desirable properties of an SQI. Let $\mathcal{D} = \{(x^{(i)}, y^{(i)})\}_{i=1}^{M}$ be a dataset of input signals $x^{(i)} \in \mathcal{X}$ and corresponding ground-truth outputs $y^{(i)} \in \mathcal{Y}$.

\subsubsection{Monotonicity}
The expected value of the given metric should be strictly increasing with respect to the SQI:  For any strictly increasing sequence $(q_n)_{n=1}^K$,
\begin{equation}
\begin{aligned}
\bar{h}_n
&:=
\mathbb{E}_{\mathcal{D}}\!\big[h(f(x), y) \mid q_n \le q(x) < q_{n+1} \big],
\\
\bar{h}_n
&<
\bar{h}_{n+1}.
\end{aligned}
\label{monotonicity_intervals}
\end{equation}

\subsubsection{Separation Margin Maximization}
In practice, SQIs are used to detect noisy samples by thresholding. Therefore, for a given threshold $\tau \in \mathbb{R}$, we define the separation margin as 
\begin{equation}
\begin{aligned}
\bar{h}_{\ge\tau} &:= \mathbb{E}_{\mathcal{D}}\!\big[h(f(x), y) \mid q(x) \ge \tau \big],\\
\bar{h}_{<\tau} &:= \mathbb{E}_{\mathcal{D}}\!\big[h(f(x), y) \mid q(x) < \tau \big],\\
\Delta(\tau) &:= \bar{h}_{\ge\tau} - \bar{h}_{<\tau}.
\end{aligned}
\label{sep_margin}
\end{equation}
SQIs should aim to maximize the separation margin achieved by the optimal threshold $\Delta^* = \max_{\tau \in \mathbb{R}} \Delta(\tau)$. This margin measures the discriminative power of the SQI with respect to the task and metric.

Importantly, both properties involve the expectations' relative ordering, rather than their absolute values. SQIs should therefore be able to detect relative performance degradations, not guarantee correct outputs.

\section{The Perturbation-Based Signal Quality Index}
\label{sec:psqi}
Motivated by the requirements in Section~\ref{sec:formalizing_sqis}, we propose a task- and metric-specific SQI based on the sensitivity of an algorithm and metric to bounded input perturbations. We hypothesize that signals of low quality for the given task and metric are affected more heavily by perturbations than high quality signals. In this section, we explain our design choices and the optimization process for the input perturbations to maximize their discriminative power for medical time-series.

\subsection{Worst-Case Metric Under Perturbations}
Let $p_{\theta}: \mathcal{X} \rightarrow \mathcal{X}$ be the perturbation function parametrized by $\theta$. We optimize $\theta$ for each input signal $x$ to find the worst-case perturbation with respect to the metric under a minimal signal-to-noise ratio (SNR) constraint. We constrain both the global and local sample-wise SNR. While the definition of the SNR may generally depend on the type of perturbations and input signals, we opt to define the \emph{global SNR} in this work as
\begin{equation}
\mathrm{SNR}(x, p_{\theta}(x)) = \frac{\|x^{\mathrm{filt}}\|_2^2}{\|p_{\theta}(x)\!-\!x\|_2^2},
\label{snr}
\end{equation}
where $x^{\mathrm{filt}}$ is the input filtered to only retain components with useful signal energy. In this work, we implement the filter as a sixth-order zero-phase Butterworth $0.5 \mathrm{Hz}$ high-pass filter to remove constant offsets and baseline wander. The \emph{local SNR} is defined as the sample-wise version of~(\ref{snr}). The sample-wise SNR constraint ensures that the allowed noise power cannot simply be concentrated on a few samples to completely mask them out. This leads to the following optimization problem for the perturbation SQI (pSQI):
\begin{equation}
\begin{aligned}
q(x) &= \min_{\theta \in \Theta}h\big(f(p_{\theta}(x)), f(x)\big), \\ 
\mathrm{s.t.} \: \frac{\|x^{\mathrm{filt}}\|_2^2}{\|p_{\theta}(x)\!-\!x\|_2^2} &\geq \gamma, \: \frac{\|x^{\mathrm{filt}}_i\|_2^2}{\|p_{\theta}(x)_i\!-\!x_i\|_2^2} \geq \beta,\ \forall i \in [N],
\end{aligned}
\label{wc_metric}
\end{equation}
where $\gamma \in \mathbb{R}^+$ is the minimal \emph{global SNR} level, $\beta \in \mathbb{R}^+$ is the minimal \emph{local SNR} level, and the parameter space $\Theta$ defines the admissible perturbations. As neither the metric nor the algorithm are guaranteed to be differentiable, we solve this optimization problem with the covariance matrix adaptation evolution strategy (CMA-ES)~\cite{hansen2001completely, hansen2019pycma}.

\subsection{Perturbation Model}
We consider additive perturbations that are parametrized as colored Gaussian noise obtained by filtering a standard Gaussian random signal through a zero-phase bandpass filter. Given a fixed noise sample $z \sim \mathcal{N}(\mathbf{0}, I_N)$, and a forward--backward (zero-phase) sixth-order Butterworth bandpass filter $\mathcal{B}_{\theta}$ with lower and upper cutoff frequencies $\theta = (f_{\mathrm{low}}, f_{\mathrm{high}})$, the perturbation function is defined as
\begin{equation}
p_{\theta}(x) = x + \frac{\|x^{\mathrm{filt}}\|_2}{\sqrt{\gamma} \, \|\mathcal{B}_{\theta}(z)\|_2} \, \mathcal{B}_{\theta}(z),
\label{additive_perturbation}
\end{equation}
which maximizes the noise power while satisfying the \emph{global SNR} constraint. We furthermore clip each sample in the perturbed signal to a maximum deviation of $(\sqrt{\beta})^{-1}$ relative to the original sample, which satisfies the \emph{local SNR} constraint. The perturbation model contains only two parameters to strike a balance between its expressivity and the difficulty of the zero-order optimization problem.

\section{Empirical Evaluations}
\label{sec:empirical_evaluations}
We evaluate the proposed pSQI on three benchmarks: ECG R-peak detection, ECG-based AF classification, and PPG-based AF classification. To ensure a fair comparison to current SQIs that are not task- and metric-specific, we specialize them for the desired task and metric with the help of a labeled dataset. Specifically, for feature-based SQIs (fSQI), we train a regressor/classifier on the labeled dataset to predict the ground-truth metrics from the features. For deep learning-based SQIs (dSQI), we train a neural network to predict the ground-truth metrics from the input signals directly. We finally evaluate the generalizability of the trained models on a separate dataset.

\subsection{Model Training and Hyperparameters}

The regressor/classifier for the fSQI is implemented using XGBoost~\cite{chen2016xgboost}. The hyperparameters \emph{max\_depth} and \emph{n\_estimators} are optimized with five-fold cross-validation and a grid search over $(5, 7, 9, 11)$ and $(100, 500, 1000)$ respectively. The neural network for the dSQI is initialized as the ECGFounder model, a state of the art ECG-based classifier~\cite{li2024electrocardiogram}. The last layer of the ECGFounder model is replaced with a new regression/classification head and the network is trained for $50$ epochs using AdamW~\cite{loshchilovdecoupled} with a batch size of $64$, a learning rate of $5 \times 10^{-6}$, and otherwise default hyperparameters. The labeled dataset is randomly split into a training ($80\%$) and validation ($20\%$) dataset and early stopping with a patience of 5 epochs is used. 

The following features from previous works~\cite{clifford2011signal,nemati2010data, orphanidou2014signal} are used in the fSQI: The kurtosis~\cite{clifford2011signal}, the skewness~\cite{clifford2011signal}, the ratio of power $\mathrm{P}(5\!-\!20\mathrm{Hz})$ to the total power~\cite{clifford2011signal}, the spectral purity~\cite{nemati2010data}, the median heart rate~\cite{orphanidou2014signal}, the shortest and longest beat-to-beat interval~\cite{orphanidou2014signal}, the mean correlation of each beat with the mean template~\cite{orphanidou2014signal}, and the percentage of matched beats detected by two different peak detectors independently~\cite{clifford2011signal}, namely the \emph{FastNVG}~\cite{emrich2023accelerated} and \emph{Neurokit}~\cite{Makowski2021neurokit} peak detectors for ECGs and the \emph{Elgendi}~\cite{elgendi2013systolic} and \emph{Charlton}~\cite{charlton2025msptdfast} peak detectors for PPGs. 

The minimal \emph{global SNR} for the pSQI is set to $\gamma = 25 \mathrm{dB}$ and the minimal \emph{local SNR} is set to $\beta = 10 \mathrm{dB}$. The CMA-ES algorithm's population size is set to 5 with a maximum of 2 iterations. We ensure that the expectations $\bar{h}_{\ge\tau}$ and $\bar{h}_{<\tau}$ in~(\ref{sep_margin}) are computed with at least 5 samples each.

\subsection{R-peak Detection}
The Glasgow University database (GUDb) is used for the first benchmark~\cite{howell2018high}. It contains cheststrap ECGs with sample accurate R-peak annotations from 25 subjects. Each subject was recorded performing 5 different tasks for two minutes. This benchmark's algorithm is the FastNVG peak detector and the metric is the $\mathrm{F}_1$ score with a tolerance of $20 \: \mathrm{ms}$~\cite{emrich2023accelerated}. The signals are segmented into $10 \: \mathrm{s}$ windows and processed independently. The fSQI and dSQI are trained on the Pulse Transit Time PPG dataset containing 22 healthy subjects performing 3 physical activities and sample accurate R-peak annotations as well~\cite{PhysioNet-pulse-transit-time-ppg-1.1.0, goldberger2000physiobank}. 

\begin{figure}[htbp]
\includegraphics[width=8.8cm]{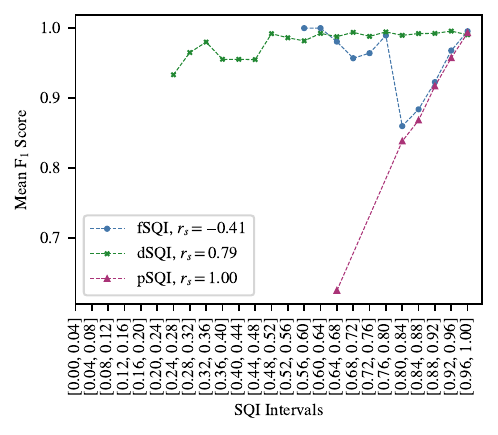}
\caption{Monotonicity evaluation of the SQIs. Each point represents the mean  $\mathrm{F}_1$ score over all samples with a SQI value in the respective interval. The Spearman correlation coefficient for each SQI is included in the legend.}
\label{bin_plot}
\end{figure}

The SQIs' monotonicity is evaluated with 25 uniform bins over the $(0, 1)$ range and quantified with the Spearman correlation coefficient. The results are shown in Fig.~\ref{bin_plot}. The mean $\mathrm{F}_1$ score is strictly increasing with respect to the pSQI, while the dSQI and fSQI have Spearman correlation coefficients of $0.79$ and $-0.41$ respectively. The pSQI furthermore has the best optimal separation margin $\Delta^* = 0.185$ in comparison to the dSQI and fSQI with values of $0.035$ and $0.033$ respectively, as seen in Table~\ref{sep_margins_table}.

\begin{table}[htbp]
\caption{Separation margin $\Delta^*$ for all SQIs and benchmarks. Bold values indicate the best performance for each benchmark.}
\begin{center}
\begin{tabular}{|c|c|c|c|}
\hline
Benchmarks            & fSQI & dSQI & pSQI \\ \hline
R-Peak Detection      & 0.033 & 0.035 & \textbf{0.185} \\ \hline
ECG AF Classification & 0.023 & 0.086 & \textbf{0.369} \\ \hline
PPG AF Classification & -0.038 & 0.108 & \textbf{0.225} \\ \hline
\end{tabular}
\label{sep_margins_table}
\end{center}
\end{table}

\begin{figure}[htbp]
\includegraphics[width=8.8cm]{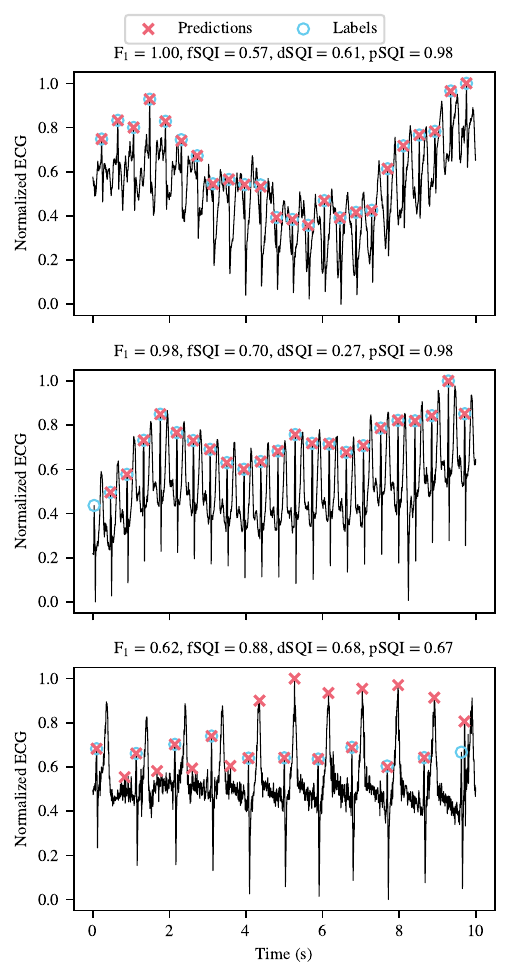}
\caption{Lowest rated ECG for each peak detection SQI. The peak predictions and labels are shown as red crosses and blue circles respectively.}
\label{worst_ecgs_peak_detection}
\end{figure}

Each SQI's lowest rated sample can be seen in Fig.~\ref{worst_ecgs_peak_detection}. Interestingly, we observe that the lowest rated pSQI sample has clear R-peaks, that would be easy to detect for a human expert. However, the FastNVG algorithm did not perform well on this sample with an $\mathrm{F}_1$ score of $0.62$ due to falsely classifying T waves and samples in the TP segments as additional R-peaks. The pSQI therefore correctly predicted the FastNVG algorithm's incorrect outputs on this sample and rated it as a signal of bad quality \textbf{for this benchmark's algorithm and metric}. In contrast, the lowest rated fSQI and dSQI samples had correct peak predictions, and were correctly rated by the pSQI as high quality signals. The fSQI and dSQI therefore did not specialize well to this benchmark's algorithm and metric.

\subsection{AF classification from the ECG}
The MIT-BIH Atrial Fibrillation Database (MIT-BIH AF) is used for the second benchmark~\cite{moody1983new, goldberger2000physiobank}. It contains 25 ten hours long ambulatory ECG recordings of subjects with AF. Each recording has two ECG leads with AF annotations, from which we use the first for our experiments. This benchmark's algorithm is the ECGFounder model and the metric is the segment-wise accuracy score. The signals are preprocessed and segmented into $10 \: \mathrm{s}$ windows according to the ECGFounder model's preprocessing pipeline~\cite{li2024electrocardiogram}. Each segment has one AF label and therefore an accuracy of either $0$ or $1$. The fSQI and dSQI are trained on the PhysioNet/Computing in Cardiology Challenge 2017 dataset containing 8528 single lead ECG recordings with AF labels lasting from $9 \: \mathrm{s}$ to just over $60 \: \mathrm{s}$~\cite{clifford2017af,goldberger2000physiobank}. In this case, due to the binary accuracy metric, the optimal separation margin is
\begin{equation}
\Delta^* \!=\! \mathbb{E}_{\mathcal{D}}\!\big[\mathbbm{1}(\hat{y} \!=\! y) \mid q(x)\!=\!1 \big] \!-\! \mathbb{E}_{\mathcal{D}}\!\big[\mathbbm{1}(\hat{y} \!=\! y) \mid q(x)\!=\!0 \big],
\label{binary_sep_margin}
\end{equation}
where $\hat{y}$ is the ECGFounder model's AF prediction. The monotonicity requirement is thereby equivalent to $\Delta^* > 0$. All SQIs fulfill $\Delta^* > 0$, however the pSQI significantly outperforms the fSQI and dSQI with $\Delta^*\!=\!0.981\!-\!0.612\!=\!0.369$ in comparison to $0.023$ and $0.086$ respectively, as seen in Table~\ref{sep_margins_table}. 

\subsection{AF classification from the PPG}
The Deepbeat test subset is used for the third benchmark~\cite{torres2020multi}. It contains 17617 $25 \: \mathrm{s}$ long single lead PPG recordings of subjects with AF. 4230 of these recordings are labeled as containing an AF rhythm. This benchmark's algorithm is the SiamAF model, a siamese network capable of classifying AF from PPG signals~\cite{guo2023siamaf}. The metric is the segment-wise accuracy score. The signals are preprocessed and zero-padded to $30 \: \mathrm{s}$ according to the SiamAF model's preprocessing pipeline~\cite{guo2023siamaf}. Each recording has one AF label and therefore an accuracy of either $0$ or $1$. The fSQI and dSQI are trained on the PPG arrhythmia dataset, which contains 118217 $10 \: \mathrm{s}$ long single lead PPG recordings with AF labels~\cite{liu2022multiclass}.

Equation~(\ref{binary_sep_margin}) again defines the optimal separation margin, with $\hat{y}$ being the SiamAF model's AF prediction instead. The dSQI and pSQI fulfill $\Delta^* > 0$, while the fSQI has a negative separation margin, as seen in Table~\ref{sep_margins_table}. The pSQI significantly outperforms the fSQI and dSQI again with $\Delta^*\!=\!0.785\!-\!0.560\!=\!0.225$ in comparison to $-0.038$ and $0.108$ respectively.

\subsection{Hyperparameter Sensitivity Analysis}
\begin{figure}[htbp]
\includegraphics[width=8.7cm]{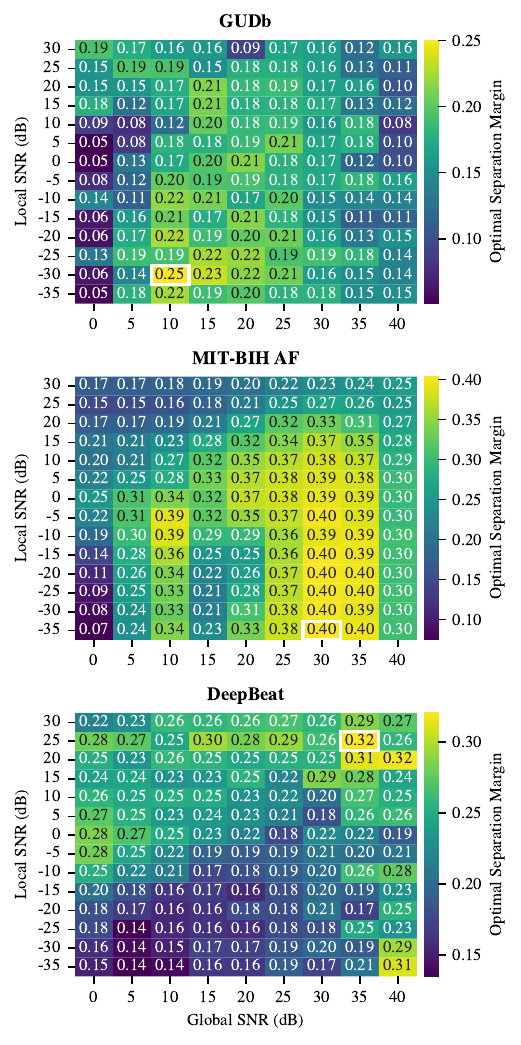}
\caption{Optimal separation margin for all benchmarks over different \emph{global and local SNR} hyperparameter values for the pSQI. The combination with the best performance for each benchmark is highlighted in white.}
\label{margin_over_snr}
\end{figure}

We additionally evaluate the pSQI's optimal separation margin for different values of the global and \emph{local SNR} hyperparameters over all benchmarks. The optimal SNR values depend on the task and input modality, as seen in Fig.~\ref{margin_over_snr}. The performance for the peak detection task is stable for \emph{global/local SNR} values between $10 \mathrm{dB}$ and $20 \mathrm{dB}$ and $-25 \mathrm{dB}$ and $-35 \mathrm{dB}$ respectively, and achieves its maximum at \emph{global/local SNRs} of $10 \mathrm{dB}$ and $-30 \mathrm{dB}$ respectively. The performance for ECG-based AF classification is stable for \emph{global/local SNR} values between $25 \mathrm{dB}$ and $35 \mathrm{dB}$ and $-35 \mathrm{dB}$ and $10 \mathrm{dB}$ respectively, and achieves its maximum at \emph{global/local SNRs} of $30 \mathrm{dB}$ and $-35 \mathrm{dB}$ respectively. Finally, the performance for PPG-based AF classification is relatively stable for \emph{global/local SNR} values between $35 \mathrm{dB}$ and $40 \mathrm{dB}$ and $20 \mathrm{dB}$ and $30 \mathrm{dB}$ respectively, and achieves its maximum at \emph{global/local SNRs} of $35 \mathrm{dB}$ and $25 \mathrm{dB}$ respectively. Consequently, stronger noise can and should be used for tasks that rely on robust ECG characteristics, such as R-peaks, while less noise is preferable for tasks that rely on finer features, such as P-waves. Furthermore, low \emph{local SNR} values are crucial for the ECG-based AF detection and peak detection benchmarks, while high \emph{local SNR} values are better for the PPG-based benchmark. Targeting local features, such as P-waves and R-peaks, with perturbations is therefore necessary to detect unreliable samples for ECGs in these benchmarks, while targeting global features such as irregular rhythms is preferable for the PPG AF benchmark. All in all, the default \emph{global/local SNRs} of 25dB and 10dB respectively result in a competitive performance for all benchmarks that could be further improved by finetuning on a labeled dataset.

\section{Conclusion}
\label{sec:conclusion}
This work introduced and formalized the notion of task- and metric-specific SQIs. Two desirable properties of SQIs were defined and used to develop objective benchmarks. This work furthermore proposed a \textbf{label-free} perturbation-based task- and metric-specific SQI. The proposed pSQI significantly outperformed alternatives that rely on labeled datasets on three benchmarks, including R-peak detection and ECG/PPG-based AF classification. 

\paragraph*{Limitations, future work, and outlook} Despite its strong empirical performance, the pSQI has limitations. The pSQI's zero-order optimization problem may result in a high computational cost due to repeated algorithm and metric evaluations. Additionally, the pSQI has only been evaluated empirically and provides no mathematical guarantees for the requirements in Section~\ref{sec:formalizing_sqis}. Finally, labeled data may still be required in practice to calculate the optimal threshold for the pSQI before using it to reject noisy samples. Future work may therefore tackle the following tasks: (a)~Extend the pSQI to solve the optimization problem in~(\ref{wc_metric}) more efficiently for differentiable algorithms and metrics, (b)~explicitly formulate the assumptions on the algorithms and metrics for which the pSQI fulfills the proposed requirements, (c)~estimate the optimal hyperparameters and threshold on one dataset and evaluate their generalizability on others.

\bibliographystyle{IEEEtran}
\bibliography{references}
\end{document}